\documentclass[a4paper,fleqn,usenatbib]{mnras}

\usepackage{txfonts}
\usepackage[T1]{fontenc}
\usepackage{ae,aecompl}
\usepackage{graphicx}	
\usepackage{amssymb}	
\usepackage{soul}

\graphicspath{ {Figures/} }

\title[Modelling a Disc-Disc Interaction]{Evidence of a {past} disc-disc encounter: HV and DO Tau}
\author[Winter, Booth \& Clarke]{
Andrew J. Winter, Richard A. Booth, Cathie J. Clarke\\
Institute of Astronomy, University of Cambridge, Madingley Road, Cambridge CB3 0HA, UK}
\date{}

\date{Accepted XXX. Received YYY; in original form ZZZ}

\pubyear{2018}

\begin{document}
\label{firstpage}
\pagerange{\pageref{firstpage}--\pageref{lastpage}}
\maketitle
\begin{abstract}
Theory and observations suggest that star formation occurs hierarchically due to the fragmentation of giant molecular clouds. In this case we would expect substructure and enhanced stellar multiplicity in the primordial cluster. This substructure is expected to decay quickly in most environments, however historic stellar encounters might leave imprints in a protoplanetary disc (PPD) population. In a low density environment such as Taurus, tidal tails from violent star-disc or disc-disc encounters might be preserved over time-scales sufficient to be observed. In this work, we investigate the possibility that just such an event occured between HV Tau C (itself a component of a triple system) and DO Tau $\sim 0.1$~Myr ago, as evidenced by an apparent `bridge' structure evident in the $160$~$\mu$m emission. By modelling the encounter using smoothed particle hydrodynamics (SPH) we reproduce the main features of the observed extended structure {(`V'-shaped emission pointing west of HV Tau and a tail-like structure extending east of DO Tau).} We suggest that HV Tau and DO Tau formed together in a quadruple system on a scale of $\sim 5000$~au ($0.025$~pc). 
\end{abstract}


\section{Introduction}

Star formation occurs predominantly in clustered environments from giant molecular clouds \citep[GMCs;][]{2003ARA&A..41...57L}. Simulations suggest that stars form in hierarchical fragmentation of these molecular clouds, resulting in small subclusters \citep[e.g.][]{2003MNRAS.343..413B}. Such subclusters interact dynamically, merging or dispersing over a similar time scale to the star formation \citep{2010MNRAS.407.1098A, 2011MNRAS.415.1967A}. In this scenario, substructure within a cluster is only directly observable over short time-scales. However, enhanced local stellar density in turn increases the chance of a close encounter between young stars \citep{2013ApJ...769..150C}, which can have significant consequences for the evolution of a circumstellar disc \citep[e.g.][]{1999MNRAS.304..425A}.

The Taurus star forming region contains almost exclusively young stars of age $\lesssim 3$~Myr and is considered an archetype of low-mass star formation, with a low stellar density and long dynamical time  \citep{Bal99}. \citet{1995MNRAS.272..213L} and \citet{2008ApJ...686L.111K} find evidence for hierarchical structure in Taurus on large scales, but not on smaller scales ($\sim0.04$~pc), and it is hypothesised that structure has been erased by dynamical interactions in this regime. Although star-disc encounters are rare in most young cluster environments \citep[e.g.][]{Win18b}, if this substructure in Taurus did indeed exist in the past then enhanced numbers of early close encounters could leave evidence in the form of truncated discs or tidal tails \citep[e.g. RW Aurigae,][]{Cab06, Dai15}. The low stellar density in Taurus also means that there are fewer disrupting influences, and any tidal tails produced in historic interactions may be preserved for periods long enough to be observed.

Photometric observations of HV and DO Tau, which have a present day separation of $90.8''$ ($0.06$~pc), by \citet{2013ApJ...776...21H} using the Photodetector Array Camera and Spectrometer (PACS) of the \textit{Herschel Space Observatory} were made at $70$~$\mu$m, $100$~$\mu$m and $160$~$\mu$m (Figure \ref{herschel}). The extended emission from each component, HV and DO, is directed towards the other, with a common envelope or `bridge' (i.e. emission connecting the two) visible at $160$~$\mu$m. While imaged at low resolution, the structure observed is reminiscent of tidal tail structures found in simulations of close encounters between disc-hosting stars \citep{1993MNRAS.261..190C, 2015MNRAS.446.2010M}. 

The following is an investigation of the hypothesis that DO Tau plus the 3 stars comprising HV Tau were originally formed as a bound hierarchical multiple, and that the present morphology of the system can be explained in terms of a close, disc mediated encounter and subsequent ejection of DO Tau from the system.  We aim to replicate observations using hydrodynamical modelling in order to understand the nature of such an interaction in terms of the disc geometry and stellar kinematics.

\section{Observational Constraints}
\label{sec:obs}

\subsection{Stellar Components}

HV Tau is a young triple system in Taurus. It is comprised of a tight optically bright binary AB, projected separation $10$~au \citep{Sim96}, and a third star HV Tau C at approximately $550$~au separation with common proper motion \citep{Duc10}. The tight binary has an estimated age $2$~Myr and a combined  mass of $\sim0.6M_\odot$ \citep{Whi01}. The separation of AB could be larger than $10$~au due to orbital eccentricity or deprojection, as suggested by a comparatively long orbital period \citep{Duc10}. A mass of $0.5-1M_\odot$  is inferred from the CO maps of the edge on disc of HV Tau C \citep{Duc10}. It is observed to be exceptionally red, with a high accretion rate \citep{Woi98, Mon00}.

DO Tau is a G star located at a projected distance $1.26 \times 10^{4}$~au ($90.8$'' at $140$~pc) west of HV Tau, which has position angle $95.3^{\circ}$ relative to DO. Mass and age estimates range between $0.3M_\odot$, $0.16$~Myr \citep{Har95} and $0.7M_\odot$, $0.6$~Myr \citep{Bec90}. The whole system is depicted with the components labelled in Figure \ref{fig:schem}.

\subsection{Disc Properties}

\begin{figure}
\centering
\includegraphics[width=0.5\textwidth]{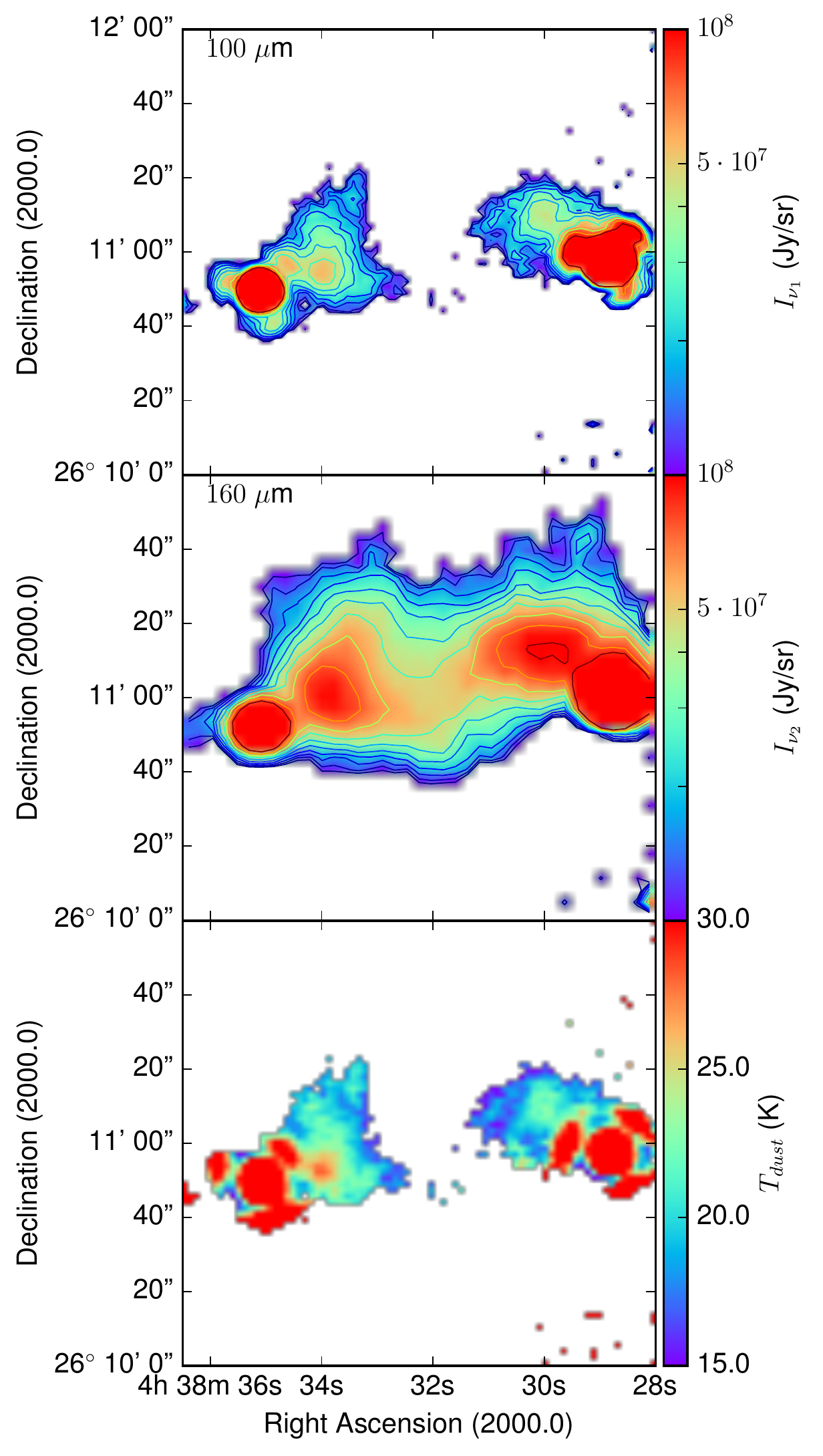}
\caption{\label{herschel} Produced using the data discussed in \citet{2013ApJ...776...21H}. The top two images are the specific intensity in the $100$~$\mu$m, and $160$~$\mu$m overlaid with logarithmic contours. Both stars appear to be associated with extended emission. The edge of the image is close to DO Tau (east), which results in excess noise. The bottom panel is the inferred dust temperature distribution assuming that the cloud is optically thin, likely yielding an overestimate close to the stars. The point spread function (PSF) in the $100$~$\mu$m observations also lead to noise in the temperature determination in these regions.}
\end{figure}

\citet{2015ApJ...808..102K} used CARMA observations and models to deduce properties of $6$ protoplanetary discs, including DO Tau. Their models found an outer disc radius of $\sim75$~au and consistent values for mass $M_{\mathrm{disc}} \approx 0.013 \, M_\odot$, inclination $\sim -33^{\circ}$, and position angle $\sim 90^{\circ}$, following the convention as described by \citet{2007A&A...467..163P}.  There remains ambiguity as to which side of the disc is closer to the observer as the quoted negative inclination angle can produce two rotation senses with the same aspect ratio. 

HV Tau A and B have no associated infrared excess and therefore are not expected to host a substantial disc, while C has an edge on disc of radius $50$~au and mass $\sim 2 \times 10^{-3} \, M_\odot$ \citep{Woi98, Sta03}.  \citet{Mon00} find that the observed disc radius does not depend on wavelength. This suggests the disc has been truncated, as otherwise the grain size-dependent radial drift of dust particles leads to a wavelength-dependent disc extent. To the contrary they note that the ratio of disc size to projected separation between C and close binary AB is $R_{\mathrm{disc}}/x_{\mathrm{min}} \equiv R_{\mathrm{tidal}} \sim 0.1$, where $R_{\mathrm{disc}}$ ($= 50$~au) is the outer disc radius, and $x_{\mathrm{min}}$ is the closest approach distance. This makes truncation due to tertiary interaction at the current separation unlikely as a ratio of around $R_{\mathrm{tidal}} \approx 0.35$ is expected if the masses of C and combined AB are equal \mbox{\citep{1999MNRAS.304..425A}}. {It remains possible that the orbit of AB is highly eccentric, and that the periastron distance is sufficiently small to cause tidally induced truncation. Alternatively, an historic encounter may have left the disc truncated.}

In modelling the disc around HV Tau C, \citet{Duc10} find an inclination $\theta_i \approx 80^{\circ}$ and PA of approximately $20^{\circ}$, corresponding to an orientation such that the blue shifted side of the disc is pointing east with the northern side closer to us. It is further noted that the coplanarity of the centre of mass of AB and the disc of C is unlikely as the nearly edge on angle would lead to a very large actual separation. \citet{Duc10} also suggest that scattered light images might imply a disc size greater than $50$~au, and gas emission alone suggests a radius up to $100$~au. A model with temperature profile $T\propto R^{-q}$ is found to fit well with $0.4<q<0.6$ and a temperature at $50$~au of $15$-$30$~K. 

\subsection{HST and Herschel/PACS Images}

\begin{figure}
\centering
\includegraphics[width=0.5\textwidth]{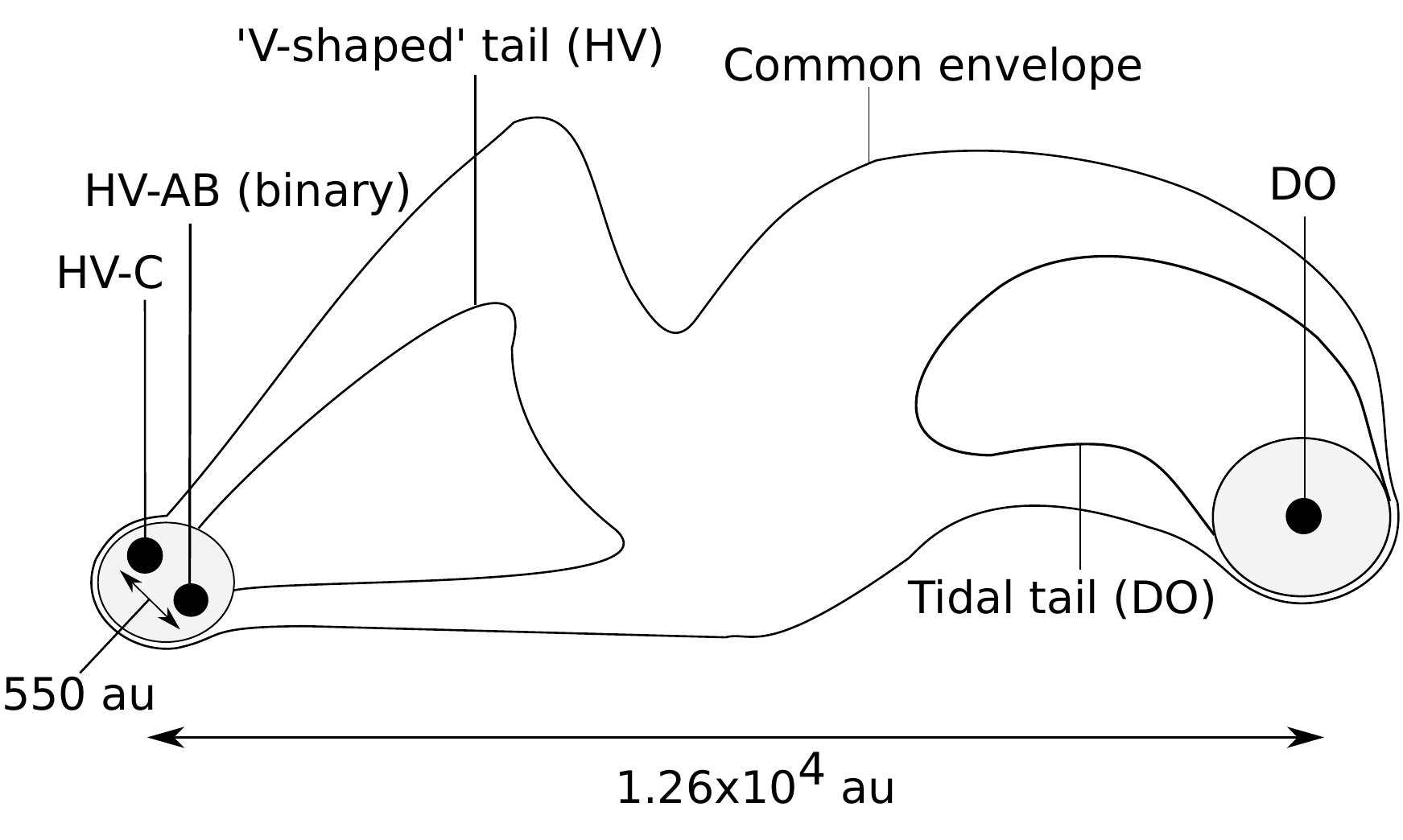}
\caption{\label{fig:schem}Schematic diagram of the $160$~$\mu$m dust emission structure visible in Figure \ref{herschel} with positions of the stellar components overlaid. The diagram is simplified to highlight the features which we aim to reproduce in our models. HV Tau is a system of  three stars, the tight binary HV-AB shown here as one point has a projected separation of $\sim 10$~au. HV-C has a PA of $\sim 45^\circ$ with respect to HV-AB, and HV has a PA of $95.3^\circ$ with respect to DO.}
\end{figure}

The Herschel/PACS survey observations of HV/DO Tau are discussed by \citet{2013ApJ...776...21H}, and we use that data to produce Figure \ref{herschel}. At $160$~$\mu$m the extended emission connects HV and DO in a common envelope. Of particular interest is the `V-shaped' emission close to HV Tau and the tail to the North-East of DO Tau (see Figure \ref{fig:schem}), seen clearly at $100$ and $160$~$\mu$m, which we aim to reproduce as the result of a disc-disc interaction producing two tidal tails. 

It has been shown in numerous studies that two tails, or a `bridge' and an external arc, can be produced as a result of prograde or inclined encounters \citep{1972ApJ...178..623T, 1993MNRAS.261..190C, 2015MNRAS.446.2010M}. Observed morphology is dependent on viewing angle and interaction parameters. Angular momentum transfer between star and disc, and therefore the quantity of circumstellar material ejected during an encounter, is a strong function of the closest approach distance \citep{Ost94, Win18}. As we will discuss in Section \ref{sec:massest}, we expect a collision between the discs, as opposed to a distant encounter, is required to produce the observed emission.

\subsection{Cloud Temperature and Mass}
\label{sec:massest}

\begin{figure}
\centering
\includegraphics[width=0.5\textwidth]{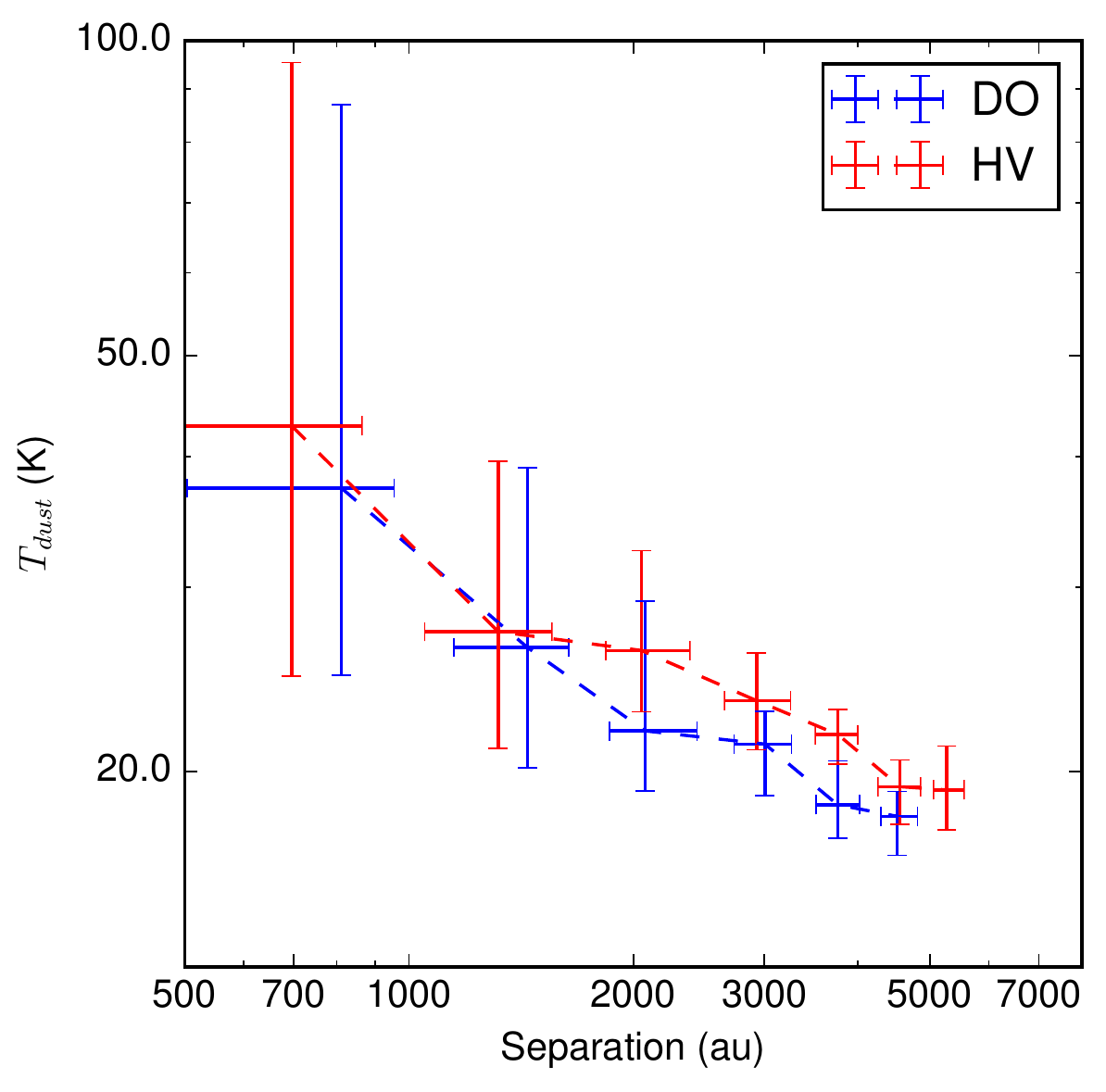}
\caption{\label{fig:tdust} Distribution of the dust temperature of each pixel in Figure \ref{herschel} as a function of separation from HV Tau (red) and DO Tau (blue). The error bars are the $1\sigma$ range in separation and temperature for a given bin of pixels. Close to the star the optical depth and the PSF result in considerable errors in the determination of temperature. }
\end{figure}

To compare the mass in the envelope of our model to that of the observations, we reproduce the expected flux at 100~$\mu$m and 160~$\mu$m using the methods outlined by \citet{1983QJRAS..24..267H}. The specific intensity of radiation at frequency $\nu$ across the envelope can be written:
$$
I_\nu = \left( 1- e^{-\tau_\nu} \right)B_\nu (T_\mathrm{dust})
$$ where $B_\nu(T_\mathrm{dust})$ is the Planck disribution at a given dust temperature $T_\mathrm{dust}$, and $\tau_\nu$ is the optical depth of the dust. The latter can be rewritten $\tau_\nu = \kappa_\nu \Sigma_\mathrm{dust}$ if we assume that $\kappa_\nu$ is spatially uniform. 

We make estimates of dust mass and temperature by assuming that $\Sigma_\mathrm{dust}$ is sufficiently small such that the cloud is optically thin ($1- e^{-\tau_\nu}\approx \kappa_\nu \Sigma_\mathrm{dust}$). While this approximation is useful away from the stars (a posteriori we find $\Sigma_\mathrm{dust} \sim 10^{-4}$~g~cm$^{-2}$ in this region) it is likely to break down locally to HV and DO Tau where $\Sigma_\mathrm{dust}$ is large. For this reason, when we come to presenting our models and final mass estimates (Section \ref{sec:hydromodel}) we will produce an intensity map from the simulation data for comparison with observations. For the two frequencies $\nu_1 = c/100~\mu$m, $\nu_2 = c/160~\mu$m we use the opacity of spherical dust grains with radius $a$ following a power law distribution $n(a) \propto a^{-3}$ between $a_\mathrm{min}=10$~nm and $a_\mathrm{max}=1.023$~cm as computed by \mbox{\citet{Taz17}}. The models in that work are based on abundances appropriate for a circumstellar disc described by \citet{Pol94}.

The measured intensities are integrated over the normalised transmission spectra for PACS $S_{\nu_{1,2}}$:
$$
I_{\nu_{1,2}} = \frac{\int I_\nu(\nu) S_{\nu_{1,2}}(\nu) \, \mathrm{d} \nu}{\int S_{\nu_{1,2}}(\nu) \, \mathrm{d} \nu}
$$  and hence
$$
\frac{I_{\nu_1}}{I_{\nu_2}} \approx  \frac{\int B_\nu(\nu;T_\mathrm{dust}) \kappa_\nu (\nu) S_{\nu_{1}}(\nu) \, \mathrm{d} \nu}{\int B_\nu   (\nu; T_\mathrm{dust}) \kappa_\nu (\nu)S_{\nu_{2}}(\nu) \, \mathrm{d} \nu}  \cdot \frac{\int S_{\nu_{2}}(\nu) \, \mathrm{d} \nu}{\int S_{\nu_{1}}(\nu) \, \mathrm{d} \nu}.
$$ We invert this expression to estimate the temperature at each pixel. The result is shown in the bottom panel of Figure \ref{herschel}. The point spread function (PSF) of the $100$~$\mu$m observations combined with the greater optical depth result in considerable errors close to the stars. However, by plotting the pixel temperature against projected distance from the nearest star we find evidence for a temperature gradient within the cloud (as expected, Figure \ref{fig:tdust}). 

Once we have the temperature in each pixel we can determine the column
density of dust that is required to match the observed emission map. This can only be performed on regions that are optically thin, and for those in which we have detections at both $100$~$\mu$m and $160$~$\mu$m. We find a dust mass of $\sim 1 -5\times 10^{-4} \,M_\odot$, depending on assumed values of $T_\mathrm{dust}$. For a dust to gas ratio $\Sigma_\mathrm{dust}/\Sigma_\mathrm{gas} = 10^{-2}$ this yields an estimate of the total cloud mass of $M_\mathrm{cloud} \gtrsim 10^{-2} \, M_\odot$. This is greater than the total present day mass of the disc around DO Tau, and would suggest that a large fraction of the circumstellar material has been ejected into the ISM (or possibly accreted onto the stellar components) during the hypothesised past encounter. {However, if the material originates in discs, the dust to gas ratio could be enhanced} \citep[e.g.] []{Ans16} and our derived cloud mass would be an overestimate.

{Based on the relative intensity of the $100$~$\mu$m and $160$~$\mu$m emission we further find evidence that the extended structure originated in a circumstellar environment. We repeat our mass estimates with opacities calculated from an ISM dust grain distribution $n(a) \propto a^{-3.5}$}, and a maximum grain size $a_\mathrm{max}= 1$~$\mu$m \mbox{\citep[see][]{Taz17}}. {Such a calculation yields lower temperatures ($\sim 10$-$20$~K) throughout the cloud and a dust mass of $\gtrsim 5 \times 10^{-3} M_\odot$ (or a total cloud mass of $\gtrsim 0.5 M_\odot$). This total mass is extremely large, and physically unlikely given the emission is associated with the stellar components of similar mass. Further, we estimate the Jean's mass:}
$$
M_\mathrm{J} \approx 2 M_\odot \left( \frac{c_\mathrm{s}}{0.2 \, \mathrm{ km/s}} \right)^3 \sqrt{ \frac{10^3 \, \mathrm{cm}^{-3}}{n_\mathrm{H}}} 
$$ {where $n_\mathrm{H}$ is the number density of hydrogen, and the sound speed $c_\mathrm{s} \approx 0.5$~km/s for a gas with $T = 15$~K. If the total mass is $0.5$~$M_\odot$ and the volume is $\sim 10^4 \times 2\cdot 10^3 \times 2\cdot 10^3$~au$^{3}$ this yields $M_\mathrm{J} \sim 0.5 \,  M_\odot \sim M_\mathrm{cloud} $. The free-fall time-scale in this case is $t_\mathrm{ff} \sim 0.03$~Myr, which is much smaller than the age of the stars. Such a cloud could be interpreted as residual material from an initial star forming core, however it is unclear whether such material could be supported against gravitational collapse on this time-scale. In addition, this interpretation offers no clear mechanism for the formation of the apparently tidal morphology. We therefore focus on the hypothesis that the material between the two systems originated in the dics around HV-C and DO.}

\subsection{Kinematics}
\label{sec:relvels}

{The proper motions DO Tau and the (unresolved) binary AB in HV Tau are recorded in \textit{Gaia DR2}} \citep{GaiaMis_16, GaiaDR2_18, Lin18}. {DO Tau has a velocity in declination $v_{\delta \mathrm{, DO}} = -21.340 \pm 0.091$~mas/yr and in right ascension  $v_{\alpha \mathrm{, DO}} = 6.128 \pm 0.126$~mas/yr. HV Tau AB has $v_{\delta \mathrm{, HV}} = -21.783 \pm 0.171$~mas/yr and in right ascension  $v_{\alpha \mathrm{, HV}} = 4.888 \pm 0.126$~mas/yr. This yields $\Delta v_\delta  = v_{\delta \mathrm{, DO}} - v_{\delta \mathrm{, HV}} = 0.29\pm 0.17$~km/s and  $\Delta v_\alpha  = v_{\alpha \mathrm{, DO}} - v_{\alpha \mathrm{, HV}} = 0.82\pm 0.24$~km/s. If the velocity vector was anti-parallel to the position vector (i.e. the systems were moving away from each other) we would expect $\Delta v_\delta \gtrsim  0$ and $\Delta v_\alpha < 0$. However, as mentioned the HV-A and -B are unresolved and multiplicity introduces uncertainties into the center of mass velocity of HV, for which an upper bound is set by the relative velocity of the AB pair} \citep[$\sim 1.5$~km/s;][]{Duc10}. {Hence the kinematic constraints are consistent with common proper motion of the two systems. Based on the projected separation, the escape  velocity is $\sim 0.4$~km/s, and it is possible that HV and DO Tau are marginally bound or unbound. The one dimensional velocity dispersion in the Taurus region is estimated to be $\sigma_v \sim 2$-$4$~km/s, although the value is uncertain due to difficulty in establishing membership}  \citep{Ber06, Riv15}. The relative proper motion components of HV and DO, which are both considerably less than this, hint at a common origin. 

{No radial velocity measurement for either star is present in the \textit{Gaia DR2}. DO Tau is estimated to have a radial velocity of $16.04 \pm 0.17$~km/s by \citet{Ngu12}, however no such estimate exists for HV Tau. Therefore we cannot place constraints on the geometry of the system using the radial velocity differential.}

\subsection{Summary of Observational Constraints}

We identify the following key criteria to consider in addressing the possibility of a previous tidal encounter.

\begin{itemize}
\item For any given parameters of a proposed fly-by, the time of the interaction should not be older than the age of the stars. Because our hypothesis requires that the stars are coeval, we already assume considerable error in the claimed ages. However, $0.16$~Myr is the lowest age estimate for any of the stellar components, and so any interaction time-scale smaller than this is feasible. Longer time-scales may also be reasonable if this is an underestimate of the age of DO Tau.
\item Disc orientations should be approximately consistent with the observations, although we note that modelling the evolution of a violent encounter over a long period of time introduces considerable uncertainty in obtaining present day orientation. To obtain a feasible solution we are motivated to explore solutions for which the disc around HV Tau C is edge on, with the plane of the disc aligned with the extended emission, while the disc around DO Tau is face on.
\item Solutions for the stellar kinematics should be consistent with the present size of the disc around HV Tau C, and hence we do not expect to see tight binary HV Tau AB orbiting C post-interaction such that $R_{\mathrm{tidal}} >0.5$, where $R_{\mathrm{tidal}}$ is here the ratio of observed disc size ($\sim50$~au) to closest approach. The closest separation between HV Tau C and DO should not be considerably less than twice the outer radius of the disc around DO Tau - i.e. $150$~au. Although it is possible that the viscous spreading of this disc may have an impact on its present extent. 
\item When recovering a flux from the surface density distribution in a given model, the dust to gas ratio required to reproduce the same flux as in the $100$~$\mu$m and $160$~$\mu$m and initial total disc mass should be sensible, and consistent between wavelengths.
\item The parameters of such an interaction should be capable of producing common envelope surrounding both stars with the structure seen in Figure \ref{herschel}. Although it may not be possible to reproduce the structure precisely, especially if the binary HV-AB has a significant effect, the aim of the modelling process is to show that the observations can feasibly result from a disc-disc interaction. 
\end{itemize}

\section{Numerical Method} 
\label{sec:method}

The complexity of the HV/DO system is approached by dividing the problem into a kinematics study of the stellar components, and hydrodynamical modelling of star-disc and disc-disc interactions. For the hydrodynamics we apply a smoothed particle hydrodynamics (SPH) treatment of the gas particles. Its computationally expensive nature means that we cannot rely on Markov Chain Monte Carlo (MCMC) or similar statistical techniques to constrain the parameters which yield the observed structure. {A large number ($\sim 500$) of low resolution models with $10^4$ particles are explored to find a promising configurations for which ejected material approximately traces the observed structure, allowing variation in disc orientations and surface density profiles} (see Section \ref{sec:discics}). Subsequently we rerun promising models with a resolution of $10^6$ particles and refining the disc properties and viewing angles to establish a model that yields extended structure closest to observations.

\subsection{Kinematic Modelling}
\label{sec:kmod}

{The first stage in obtaining a model is exploring the kinematic parameter space of a multiple encounter of a three star system (DO, HV-C and HV-AB, the latter we will consider one star - see below) to find solutions which satisfy the dynamical conditions discussed in Section} \ref{sec:obs}. {As in the case of the hydrodynamics, we cannot use an MCMC exploration of the kinematic parameter space due to the chaotic nature of the three body problem. Instead we search for a (probably non-exhaustive) library of kinematic solutions for further hydrodynamical modelling.  We do this by uniformly varying parameters which describe the initial conditions of the three bodies and checking for consistency with observations. Viable solutions are expected to be initially bound, but we do not have further a priori constraints. We apply the following parametrisation of the problem (sampling uniformly over each within the defined range) as it allows us to minimise the size of the exploration space by choosing likely ranges, with the caveat that drawing statistical conclusions from our kinematic library is problematic.}  We simulate the trajectories of the three star particles by applying the $N$-body 4$^{\mathrm{th}}$ order Hermite integrator \citep{1992PASJ...44..141M} in the \textsc{gandalf} code  \citep[which is also used for the SPH simulations described in Section \ref{sec:hydromodel},][]{Hub18}.

\begin{figure}
\centering
\includegraphics[width=0.5\textwidth]{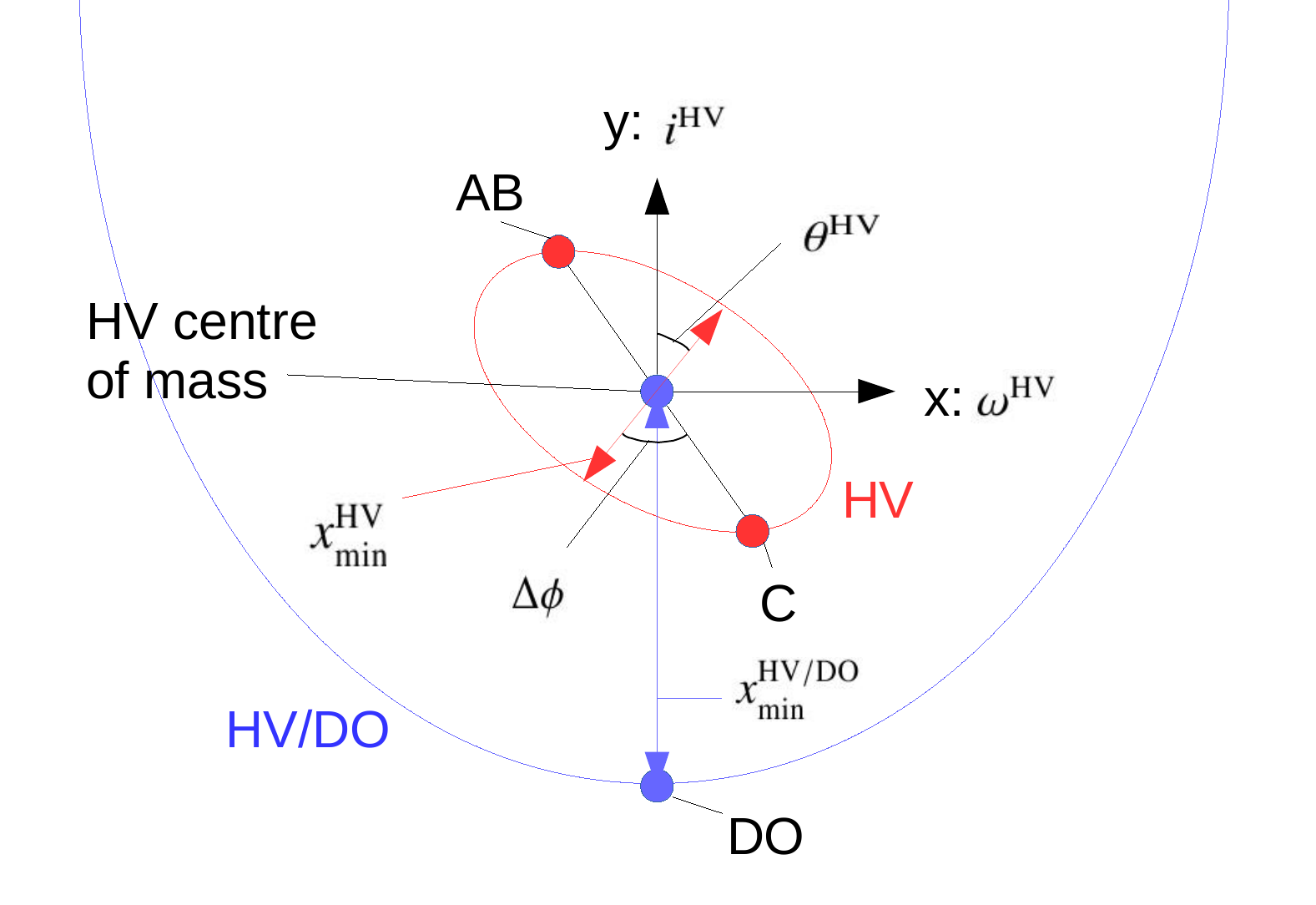}
\caption{\label{fig:icdiag} {Schematic diagram illustrating the parameters used to define the initial conditions for our 3-body simulations. The blue line traces the HV/DO trajectory, with coordinates centred on the centre of mass of the HV system. The red line traces the HV-AB/C trajectory. The circular markers represent the locations of the components of each orbit at the time of the closest approach between DO Tau and the centre of mass of HV (blue circles). The positions of HV-AB and -C are shown as red circles. The angles as discussed in the text are annotated}}
\end{figure} 

{Our parametrisation is described below, and illustrated in Figure} \ref{fig:icdiag}. Firstly, we are helped by the small separation of the binary AB, which we hereafter consider as a single star with the combined mass. With this approximation all stellar components now have the same mass within errors, and this is estimated to be $0.7 \, M_{\odot}$. In order to parametrise the interaction of the three remaining stellar components, we consider two distinct orbital equations of the form 
\begin{equation}
\label{orbit}
x = \frac{h^2}{\mu}\left( \frac 1 {1+ e \cos\left( \phi -\theta\right)}\right)
\end{equation}
for HV and for HV/DO, where HV is the orbit of HV-C and HV-AB, while HV/DO is the `two-body' system comprised DO and the centre of mass of HV. 
In Equation \ref{orbit}, $x$ is the separation between bodies, $\phi$ is phase, $\theta$ is the angle of the periastron in the plane (equivalent to rotation in the $z$-axis), $h$ is the specific angular momentum and $\mu = G(m_1 +m_2)$. For HV/DO we fix $\theta =0^\circ$. In the case of HV, the orbit of C and AB is rotated in the $y$-axis by angle $i$ and in the $x$-axis by angle $\omega$. The final parameter $\Delta \phi$ is defined as the difference in phases as DO reaches periastron, with respective separation $x$ found for the initial conditions by integrating back along the arc of both orbits. This leaves 8 initial values fully parameterising the system: $e^{\mathrm{HV/DO}}_0$, $x_{\mathrm{min}}^{\mathrm{ HV/DO}}$, $e^{\mathrm{HV}}_0$, $x_{\mathrm{min}}^{\mathrm{HV}}$, $\theta^{\mathrm{HV}}$, $i^{\mathrm{HV}}$, $\omega^{\mathrm{HV}}$, $\Delta \phi$.

\begin{table*}
\centering 
 \begin{tabular}{c c c c c c c c c} 
 \hline
 & $x_{\mathrm{min}}^\mathrm{HV/DO}$/au &  $e^{\mathrm{HV/DO}}_0$ &   $x_{\mathrm{min}}^\mathrm{HV}$/au & $e^{\mathrm{HV}}_0$ & $\theta^{\mathrm{HV}}$ /$^\circ$ & $i^{\mathrm{HV}}$ /$^\circ$ & $\omega^{\mathrm{HV}}$ /$^\circ$ & $\Delta \phi$ /$^\circ$ \\ [0.5ex] 
 \hline
Range & $0$-$2000$ &  $0$-$1$  & $100$-$1500$  & $0$-$1$ & $0$-$360$ & $0$-$360$ & $0$-$180$  & $0$-$360$ \\
 [1ex] 
 \hline
\end{tabular}
\caption{Parameter range searched for solutions to the present day arrangement of HV and DO Tau. } 
\label{table:par_range}
\end{table*}

{The ranges for each parameter over which we search for successful kinematic solutions are summarised in Table} \ref{table:par_range}. We focus on the solutions for which DO is initially bound to HV ($e^{\mathrm{HV/D0}}_0<1$) {as they offer the most likely scenarios for a close encounter between stellar components. Further, highly hyperbolic encounters in a low density stellar environment are physically unlikely.} We apply one further restriction that configurations for which the energy of the HV initial orbit exceeds the energy of the DO trajectory are discounted. This is both because in this regime our orbital parametrisation does not make physical sense, and because our investigation finds that solutions for which the orbital energies are comparable are also relatively rare. We search uniformly over the remaining parameter space for successful solutions.

Our criteria for a `successful' kinematic solution are as follows. A lower limit of $50$~au is placed on all interactions as this is a conservative constraint, a distance below which either disc would be significantly over-truncated. Additionally an upper limit on the closest approach between HV Tau C and DO Tau is set at $300$~au. This is motivated both by the present day disc outer radii and the study of \mbox{\citet{2015MNRAS.446.2010M}} and our own findings that a close flyby is required to produce the observed extended structure in the tidal tails (see Section \ref{sec:sph}). After encounter, DO must either be unbound from the whole system, or reach a maximum separation $>1.2 \times 10^4$~au. HV Tau C and AB must remain bound. Acceptable final maximum separation of HV wide binary is defined to be between $400$ and $1500$~au, in line with observed projected separation of $550$~au. A minimum periastron distance is placed at $125$~au to prevent over-truncation of the disc around HV Tau C.

\subsection{Hydrodynamics Model}
\label{sec:hydromodel}

The SPH code \textsc{gandalf} is used in the simulation of the discs \citep{Hub18}. It is adapted here to include a locally isothermal equation of state as a function of radial separation from the nearest star. Self-gravity is disregarded, the gravitational potential being dominated by the stellar component. 

Artificial viscosity parameters as prescribed by \citet{1997JCoPh.136...41M} are applied to minimise the effects of viscous diffusion in the tidal tails. However, inevitably at the required integration times on the order of $0.1$~Myr, the effect of numerically accelerated viscous spreading and magnified inter-particle torques will result in a loss of structure. This is especially the case where there is considerable mass loss from the disc, as during the violent interactions necessary to produce significant external structure. 

\subsection{Disc Interaction Initial Conditions}
\label{sec:discics}

\citet{2005ApJ...629..526P} showed that for discs in which there is significant mass transfer one cannot analogously extrapolate structure from star-disc interactions, and hence both discs are required simultaneously for all models where closest approach is of order the disc radius. For disc-disc simulations the work of \citet{2015MNRAS.446.2010M} offers a starting point in terms of the expected closest approach between HV Tau C and DO Tau, where extremely close interactions with $R_{\mathrm{tidal}}\sim 10.0$ both result in the near-destruction of the original discs and also in significant sapping of orbital energy and stellar capture (although a large disc mass approximately $10$\% of the star mass is used in this study). Conversely, encounters with a wide closest approach such that $R_{\mathrm{tidal}} < 0.5$ do not produce significant external structure.  

{Due to the uncertainty in the line of sight separation (and therefore the angle of orientation) of the present day system, the appropriate disc orientations are not immediately clear. For the initial conditions of the three star encounter, a snapshot is taken from an appropriate kinematic model at a time before close encounter. In order to ensure that discs are dynamically settled prior to the encounter, this time is chosen to be five orbital periods at the radius of the outer disc before closest approach between any two stellar components. The discs around HV Tau C and DO are added at an orientation which matches the present day orientation if the two stellar systems are in the plane of the sky. The simulation is then continued with SPH discs included to examine the hydrodynamic evolution of the multiple star interaction. Subsequently disc orientations in promising models are modified to better match the extended structure.}

The surface density profile of the discs is both important to the structure and quantity of ejected material, and hard to constrain given that it may be significantly altered in a close interaction. It is treated as a power law such that
$$
\Sigma = \Sigma_0 \left(\frac{R}{R_0}\right)^{-p}
$$ where both `shallow' ($p=0$) and `steep' ($p=1$) surface density gradients are tested.

Temperatures in the disc are defined by distance to the nearest star by
$$
\centering
T =\mathrm{max}\left\{T_0 \left( \frac R R_0\right)^{-q}, 15\mathrm{~K}\right\}
$$ with a value of $q=0.6$ and a temperature at $50$~au of $20$~K is adopted for HV Tau C and the same profile assumed for DO Tau. Variations in temperature are expected only to have a modest effect on the observed structure as a result of star-disc interaction \citep{Dai15}. Our choice of temperature profile for the hydrodynamic simulations is based on the observations by \citet{Duc10} and is lower than the observed temperature through the extended cloud discussed in Section \ref{sec:massest}. This discrepancy could be due to heating of the ejected material during the disc-disc encounter, which we do not model here as there are considerable uncertainties in the temperature estimates. The temperature in both the disc and the cloud are both empirically derived and therefore represent reasonable choices.

Outer radii of the discs prior to interaction are not well constrained, as it is unknown the proximity of the closest approach and therefore the extent of truncation by the initial fly-by. Further, the post-interaction relaxation of the disc, including viscous spreading and possible further dynamical binary interactions in the case of HV-C, is not well characterised. To eject sufficient material to produce observed structure, initial tests suggest that $R_{\mathrm{out}}$ such that $R_{\mathrm{tidal}}\equiv R_{\mathrm{disc}}/x_{\mathrm{min}}\approx 0.8$ is reasonable. This is the initial estimate for a given kinematic model, and the outer radii are subsequently tuned to fit observations. The inner radius is defined to be $R_{\mathrm{disc}}/20$. Choosing a conservative inner radius is necessary given that a significant proportion of the discs pass though each other. {The smoothing lengths of the sink particles are chosen to be half of the inner radius of the disc with the smallest extent. }

The final parameter required to define the disc interactions is the relative masses of the two discs (i.e. how many SPH particles each contains). For each configuration we allow the mass ratio to vary.

\section{Modelling Results}
\label{sec:sph}

Before presenting our chosen model, we note that while we will refer to it as the `best-fitting model', this is in the sense that it best matches observations of all the models studied. As discussed, the size of the parameter space involved and the computational expense of the simulations means that the number of models examined is not exhaustive, and that usual statistical parameter space exploration techniques were not practical.

\subsection{Kinematic Properties}

\begin{table*}
\centering 
 \begin{tabular}{c c c c c c c c} 
 \hline
 $x_{\mathrm{min}}^\mathrm{HV/DO}$/au &  $e^{\mathrm{HV/DO}}$ &   $x_{\mathrm{min}}^\mathrm{HV}$/au & $e^{\mathrm{HV}}$ & $\theta^{\mathrm{HV}}$ /$^\circ$ & $i^{\mathrm{HV}}$ /$^\circ$ & $\omega^{\mathrm{HV}}$ /$^\circ$ & $\Delta \phi$ /$^\circ$ \\ [0.5ex] 
 \hline
 $864$ &  $0.85$  & $653$  & $0.38$ & $28$ & $158$ & $10$ & $94$ \\
 [1ex] 
 \hline
\end{tabular}
\caption{Kinematic parameters of the best-fit model. Parameters are defined in Section \ref{sec:kmod}.} 
\label{table:kinparams}
\end{table*}

\begin{figure}
\centering
\includegraphics[width=0.5\textwidth]{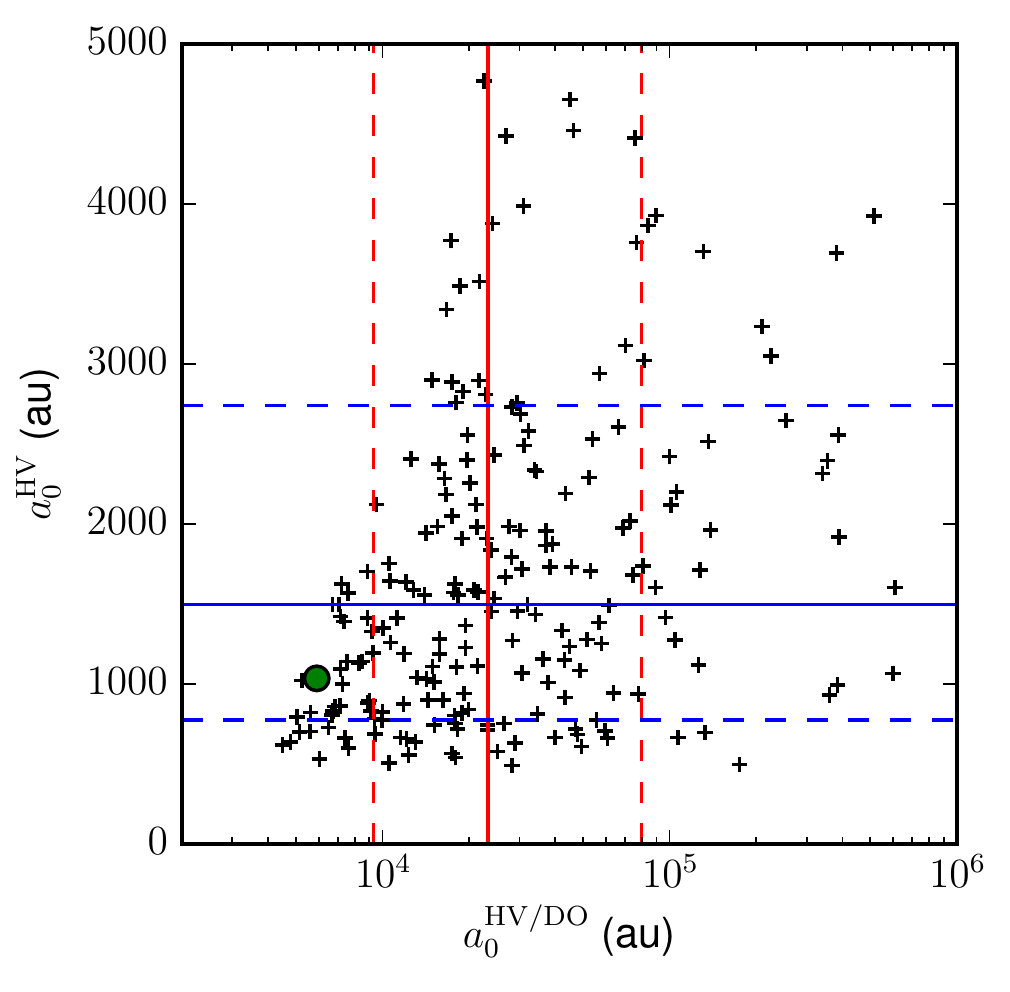}
\caption{\label{fig:kindist}  {The distribution of the initial semi-major axes of the HV ($a_0^\mathrm{HV}$) and HV/DO ($a_0^\mathrm{HV/DO}$) trajectories for successful solutions of our kinematic parameter space exploration. The solid lines (horizontal blue for $a_0^\mathrm{HV}$ and vertical red for $a_0^\mathrm{HV/DO}$)  represent the median of the results. The associated dashed lines indicate the associated $16^\mathrm{th}$ and $84^\mathrm{th}$ percentile values.  The green circle represents the location of our chosen `best-fit' solution in reproducing the extended emission between the stellar systems (see Section \ref{sec:hydrobridge}). }}
\end{figure}

{The distribution of semi-major axes in the initial systems ($a_0^\mathrm{HV/DO}$ and $a_0^\mathrm{HV}$) are shown for successful kinematic solutions is shown in Figure} \mbox{\ref{fig:kindist}}. As discussed in Section \ref{sec:kmod}, it is not possible to draw statistical conclusions from this distribution. However, we note that most solutions exist for $a_0^\mathrm{HV/DO} \sim 10^4$~au, although the model which best reproduces the extended bridge structure (Section \ref{sec:hydrobridge}) has $a_0^\mathrm{HV/DO} \approx 5800$~au. The parameters of this model are presented in Table \mbox{\ref{table:kinparams}}. {We note that the orientation of the HV/DO angular momentum vector is approximately anti-parallel that of HV-AB/C. This reversal of the orbits appears surprising. However, if the forming stars were initially separated by $\sim 4 \cdot 10^4$~au (initial apastron) it is possible that local velocity fields in the collapsing gas of the primordial system lead to non-aligned orbits.}

The important dynamical properties of the chosen kinematic model are summarised in Table \ref{table:kinprops}. By integrating backwards, all stellar components in this model are found to remain bound on time-scales $>1$~Myr. Initially HV-AB/C has an orbit with a semi-major axis $a^\mathrm{HV} _0 \approx 10^3$~au, and eccentricity $e_0 \approx 0.37$. The encounter with DO removes angular momentum from the HV system, and results in DO being marginal bound, with a large semi-major axis $a_\mathrm{f}^{\mathrm{HV/DO}} \approx 1.5 \times 10^4$~au, sufficient to reach the observed present day projected separation. 

The closest encounter between each stellar component is also consistent with observations. The single encounter between HV Tau C and DO Tau is the closest between any of the components at $285$~au, and is close enough to truncate discs to $\sim 100$~au. No interaction involving AB is close enough such that a $\sim10$~au binary is likely to be disrupted. The minimum distance between HV Tau C and AB is equivalent to the final periastron distance as no closer interaction occurred.

Finally, the time since the closest encounter to reach the projected present day separation for our preferred system orientation is $\sim 0.1$~Myr, which is consistent with even the lowest estimate for the age of any of the stellar components. 

\begin{table}
 \centering 
 \begin{tabular}{l  c  || c  c  c c} 
 \hline
 & $x_{\mathrm{min}}$/au &  $a_0$/au &  $e_0$ & $a_{\mathrm{f}}$/au &  $e_{\mathrm{f}}$\\ [0.5ex] 
 \hline
 HV-C/DO & $285$  & - & - & - & - \\ 
 HV-AB/DO & $657$ & - & - & - & -  \\
 HV-AB/C & $445$ & $1.05 \cdot 10^3$ & $0.37$ & $859$ & $0.48$\\
 HV/DO  & - & $5.76\cdot 10^3$ & $0.85$ & $1.48 \cdot 10^4$ & $0.95$   \\
 [1ex] 
 \hline
\end{tabular}
\caption{Dynamical properties of the stellar components of the best-fitting model, where $x_{\mathrm{min}}$ is the closest approach $a_0$, $a_{\mathrm{f}}$, $e_0$, $e_{\mathrm{f}}$ are the initial and final semi-major axes and eccentricities of the binaries respectively.} 
\label{table:kinprops}
\end{table}

\subsection{Disc properties}

\begin{figure*}
\centering
\includegraphics[width=\textwidth]{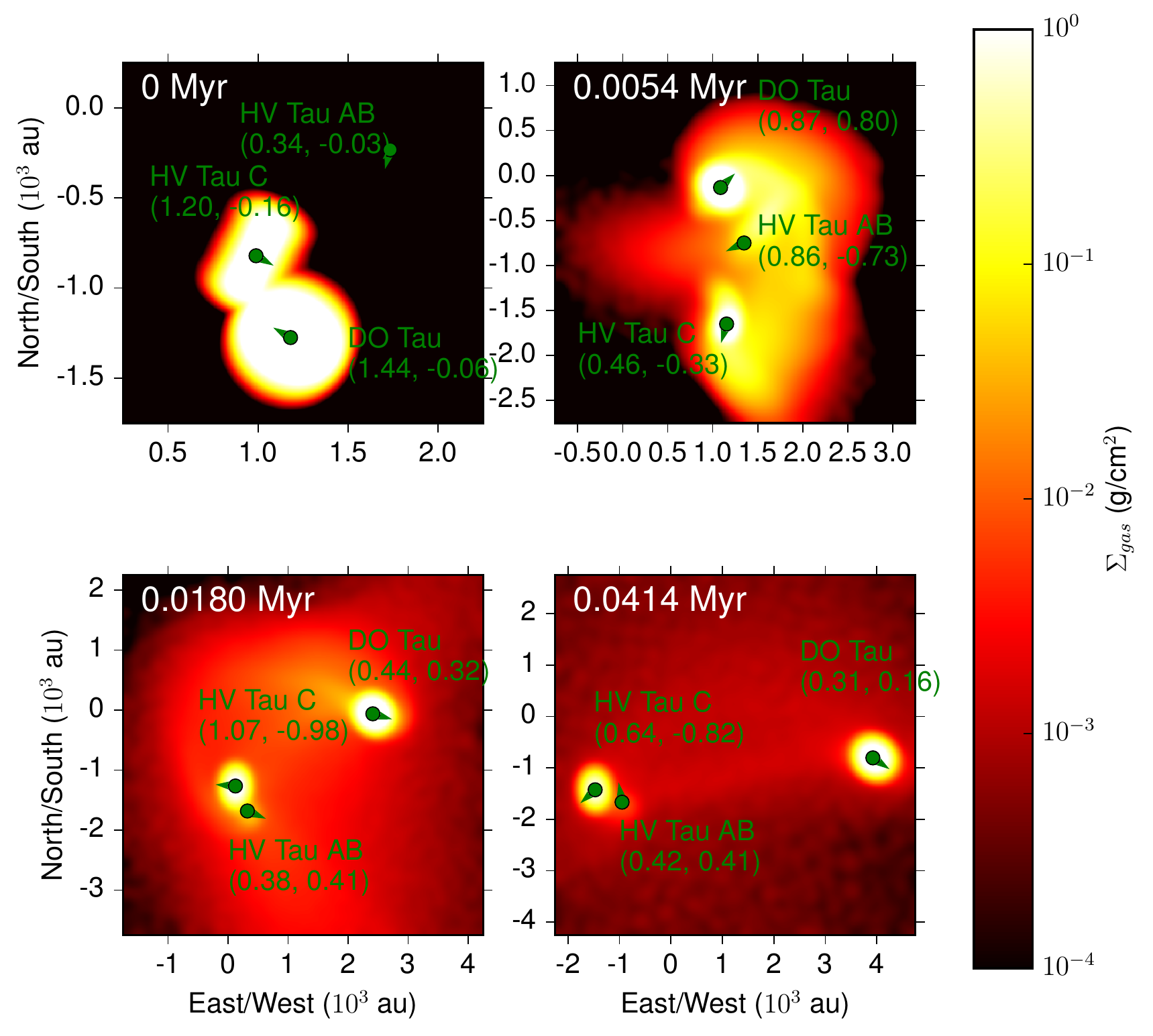}
\caption{\label{fig:ICs} Snapshots of our chosen model before and after the disc-disc interaction. The colour scale represents the gas surface density normalised to give the correct flux scale in Figure \ref{finalfig}, and the orientation is the same as in that Figure. Stellar components are marked with green circles. The numbers in brackets are the magnitude of the proper motion and the radial velocity in km/s respectively, with the direction of proper motion indicated by an arrow. HV Tau AB is considered in our models to be a single sink particle, as discussed in the text. }
\end{figure*}

The properties of the circumstellar discs found by tuning to best match the Herschel observations in Figure  \ref{herschel} are shown in Table \ref{table:discprops}, and the snapshots of the gas surface density distribution during the encounter are shown in Figure \ref{fig:ICs}. The initial radii for HV Tau C and DO Tau discs are $320$~au and $355$~au respectively, which means that the stellar components penetrate the discs at the closest approach distance of $285$~au. We find that both a smaller mass and outer radius are required for the disc around HV-C with respect to DO. The present day observed disc mass ratio is $M_{\mathrm{disc}}^{\mathrm{HV-C}}/M_{\mathrm{disc}}^{\mathrm{DO}} \approx 0.15$, while our chosen model has an initial mass ratio of $0.33$. At the time of our chosen snapshot this ratio in the simulation becomes $\sim 0.13$, with the disc around HV Tau C losing a greater fraction of the initial mass.

In our model the orientation is such that the disc around HV Tau C is approximately edge on with the plane along the direction of the `V'-shaped emission, as suggested by observations (see Figure \ref{fig:ICs}). The disc around DO Tau is also approximately face-on, and thus the geometry of the system is compatible with the observed extended structure discussed below. These disc orientations lead to a collision in which the discs collide approximately perpendicular in a strongly penetrating encounter. This violent interaction induces significant pressure gradients and justifies the need for hydrodynamic simulations.

\begin{table}
 \centering 
 \begin{tabular}{l  c  c  c  c  c  c  c  c  c  c  c} 
 \hline
 & $R_{0}$/au &  $x_{\mathrm{min}}$/au & $R_\mathrm{obs}$/au & $M_{\mathrm{rel}, \, 0}$ & $M_\mathrm{obs}$/$M_\odot$ &  $p$ \\ [0.5ex] 
 \hline
 HV-C & $320$  & $285$ & $\sim 50-100$ &  $0.33$ & $\sim0.002$ & 0 \\ 
 DO & $355$ & $285$ & $\sim 75$ &  $1.0$  & $0.013$& 0 \\
 [1ex] 
 \hline
\end{tabular}
\caption{Disc properties of the best-fit model. The quantities are as follows: $R_{ 0}$ is the initial outer radius of the disc, $x_{\mathrm{min}}$ is the closest encounter with any stellar component, $M_{\mathrm{rel},0}$ is the initial relative mass of each disc,  $M_{\mathrm{obs}}$ is the observed total disc mass, $p$ is the power law index for the surface density. The subscript $0$ pertains to initial values in the model and `obs' the observed (present-day) values.} 

\label{table:discprops}
\end{table}

\subsection{External Structure}
\label{sec:hydrobridge}
\begin{figure*}
\vspace{-10pt}
\centering
\includegraphics[width=\textwidth]{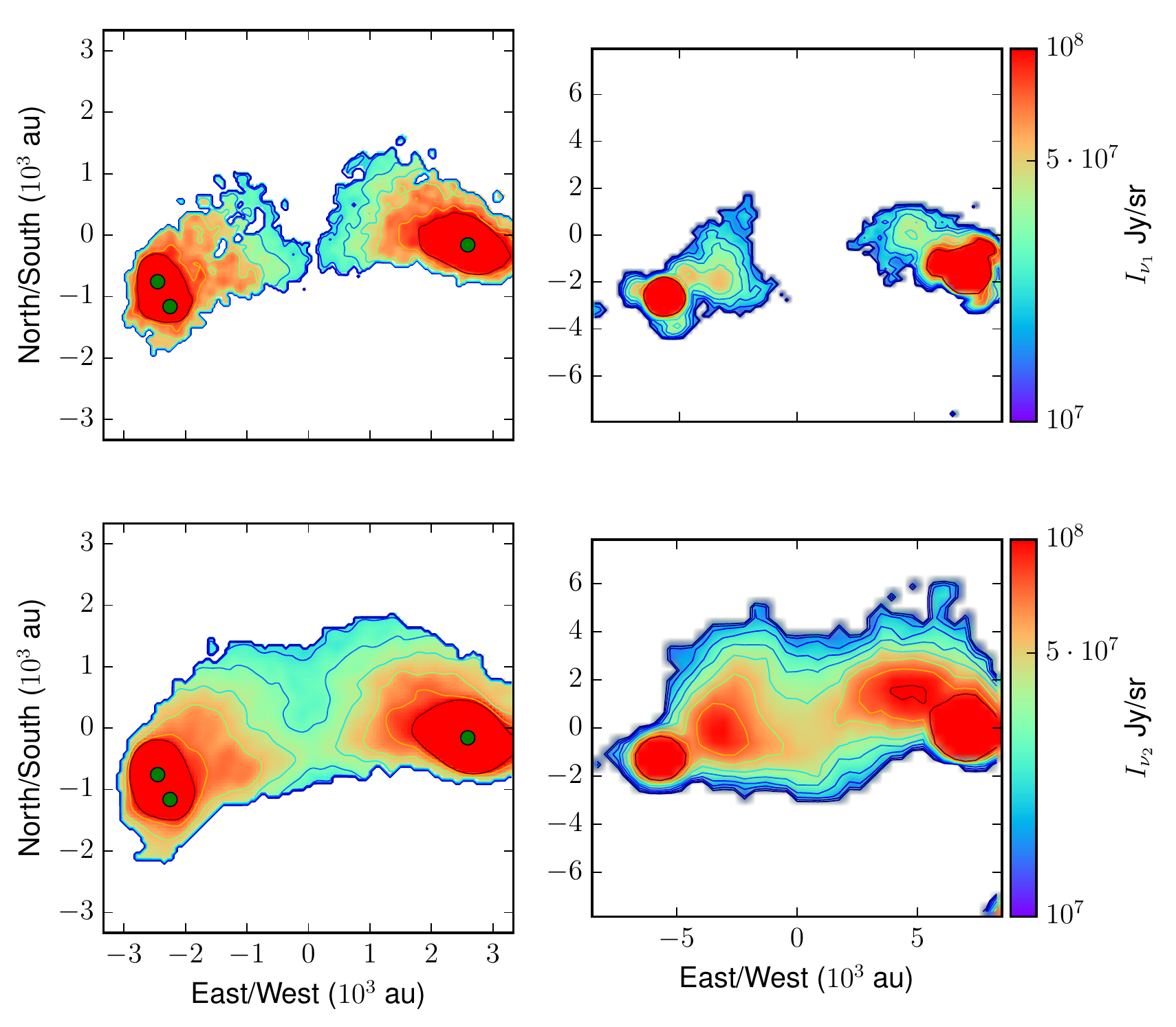}
\caption{\label{finalfig} Surface flux distribution of our chosen model (left) next to the observations (right) at $100$~$\mu$m (top) and $160$~$\mu$m (bottom). All fluxes are truncated at the $3\sigma$ background noise level in the respective wavelength observations.  The model snapshots are at $\sim 4 \times 10^4$~years after the disc-disc encounter between HV-C and DO. This is a shorter than the time required to reach the present-day separation, and is chosen due to numerical limitations (see text for details).}
\end{figure*}

\begin{figure}
\vspace{-10pt}
\centering
\includegraphics[width=0.5\textwidth]{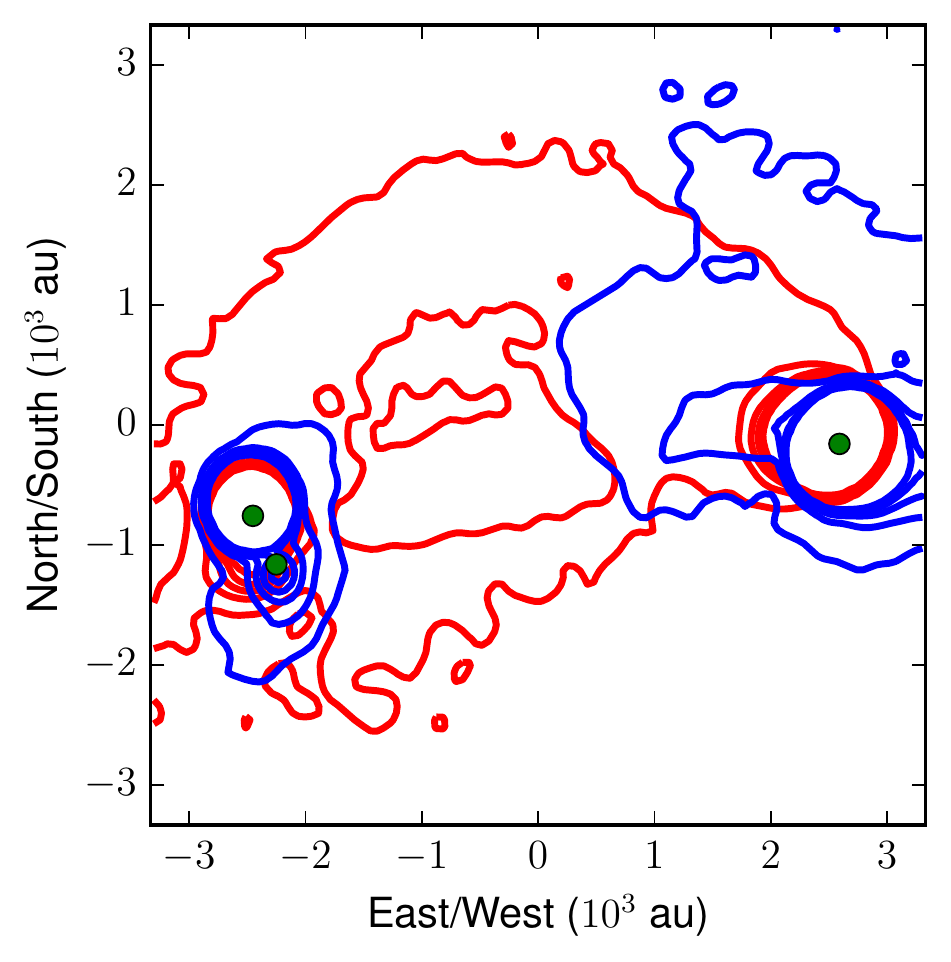} 
\caption{\label{fig:vzgas} Simulated variation in line of sight gas velocity $\delta v_z = v_z-\langle v_z \rangle $ density contours in the extended gas cloud. The contours are spaced over a factor 5 in surface density in arbitrary units. The blue contours are for SPH particles with $0$~km/s $< \delta v_z < 1$~km/s, while the red contours are for  $-1$~km/s $< \delta v_z < 0$~km/s.}
\end{figure}

In order to reproduce the extended structure between HV and DO, we have introduced a moderate temperature gradient with respect to the projected distance $d$ from each stellar component:
$$
T_\mathrm{dust} = 35 \, \mathrm{K} \left( \frac{d}{950 \, \mathrm{au}}\right)^{-0.32}
$$ with a maximum temperature of $35$~K, which is consistent with the temperature profile found in Section \ref{sec:massest}. The resulting surface brightness of the extended structure at $100$~$\mu$m and $160$~$\mu$m in our model is shown in Figure \ref{finalfig}. In order to obtain this flux distribution we have had to assume a large initial total gas mass of $M_{\mathrm{tot},0} = 0.18\, M_\odot$ (with $\Sigma_\mathrm{dust}/\Sigma_\mathrm{gas} = 10^{-2}$). This is on the order of the mass we would expect if the interaction occurred at an early evolutionary stage. Approximately $50 \%$ of the mass is accreted in our simulations at the time of the snapshot, which leaves $0.09 \, M_\odot$ total mass, of which $\sim 0.027 \, M_\odot$ is retained in the disc around DO Tau and $3.5 \times 10^{-3} \, M_\odot$ in that of HV Tau C. {The remaining mass occupies the external structure.} These disc masses are a factor $\sim 2$ greater than the present day, and indeed the mass of the total system is expected to be an {overestimate due both observational and numerical factors. First we find resolution-dependent diffusion of SPH particles into the ISM (away from what we consider the `bridge' between HV and DO). As we increase the resolution, for simulations run at a resolution lower than $10^6$ particles, a smaller fraction of SPH particles are lost to the ISM. Therefore we expect that increasing the resolution further would decrease the required total initial mass of the system. Additionally, increasing the initial radii of the discs has a similar effect of increasing the mass of the bridge while preserving the observed structure; however this additionally enhances accretion rates and therefore compounds resolution issues at late times. Alternatively, the dust-to-gas ratio in the original discs may be enhanced \citep{Ans16}, which would mean our gas mass is overestimated. }

We also note that we have chosen a snapshot at a separation between HV and DO of $\sim 5\times 10^3$~au, half of the observed present day separation. This is because, as discussed in Section \ref{sec:hydromodel}, resolution effects mean that the structure diffuses as the model is integrated in time. Integrating further to the present day results in a numerical loss of structure due to low resolution in the region between the stellar components. Contrary to the diffusive numerical effects described above, this means that additional initial mass would be required to produce sufficient surface density at the present day separation.

Overall, the main features seen in the $100$~$\mu$m and $160$~$\mu$m observations are well produced in our model, {namely the V-shaped emission close to HV Tau and the tidal tail close to DO Tau.} The broad envelope shape is less well reflected in our models, however we note that these regions have a low resolution of SPH particles which can result in a loss of structure. Additionally, uncertainties in the temperature profile discussed in Section \ref{sec:massest}, particularly at the outer edge and centre of the envelope where we only have detections at $160$~$\mu$m, mean that we are unable to accurately map the surface density to an intensity distribution. However, the agreement between our model and the observations is sufficient to suggest that a disc-disc interaction $\sim 0.1$~Myr ago is a viable mechanism by which the extended structure between HV and DO Tau has been produced. 

\subsection{Gas Velocity}

In Figure \ref{fig:vzgas} {we demonstrate that we expect to find some substructure in the line of sight gas velocities. The standard deviation in line of sight velocity of the SPH particles $v_z$ for the best fit model is $\sigma_{v_z} \approx 1.3$~km/s. We divide the deviation from the mean gas velocity $\delta v_z = v_z - \langle v_z \rangle$ into two bins, red shifted ($-1$~km/s$<\delta v_z <0$~km/s) and blue shifted ($0$~km/s$<\delta v_z <1$~km/s). The results in Figure} \ref{fig:vzgas} illustrate both the large scale velocity structure of the whole system, and the line of sight motion of the wide binary HV Tau C and AB.

Although, as previously discussed, the present day system is at approximately double the separation of the snapshot, Figure \ref{fig:vzgas} {is indicative of the velocity field we would expect to obtain from observations if a past encounter produced the observed extended emission. Future observations of the gas in the region can be compared with our results to establish the likeliness of the scenario we suggest here.}

\section{Conclusions}
\label{sec:conc}

We have used hydrodynamic modelling to lend evidence to the conclusion that the three stars making up HV Tau and the apparently unrelated star DO Tau had a past encounter $\sim 0.1$~Myr ago. While it is difficult to make hard conclusions about the nature of the dynamical history of the system and subsequent disc evolution, our modelling suggests the following scenario:

\begin{itemize}
	\item HV Tau A, B and C initially formed a quadruple system with DO Tau $\gtrsim 0.1$~Myr ago, with a spatial scale of $\sim 5000$~au (and an orbital period of $\sim 0.3$~Myr).
	\item The highly eccentric orbit of DO Tau led to a close encounter with HV Tau C $0.1$~Myr ago. During this encounter the disc around HV Tau C interacted strongly with the disc around DO Tau, leading to rapid accretion and truncation of the discs. {This was likely the first encounter and therefore we expect the age of the original system to be $\lesssim 0.4$~Myr. }
	\item Subsequent to this encounter the DO Tau trajectory became either marginally bound or marginally unbound to reach a separation $> 10^4$~au. 
	\item The tidal tails of this event can be observed in the $160$~$\mu$m dust emission to the present day. 
\end{itemize}

In terms of the history of Taurus, this supports the idea that there previously existed substructure down to smaller scales which has now been dynamically erased \mbox{\citep{2008ApJ...686L.111K}}. Given the improbability of such a close encounter producing tidal tails that can be observed for time-scales $\sim 1$~Myr after the encounter, it is likely that many more such encounters which cannot be inferred have also occurred.

\section*{Acknowledgements}

We are appreciative for the comments of the anonymous referee which helped to significantly strengthen the arguments presented here. We would like to thank Marco Tazzari for useful discussion and for providing the dust opacities used in this work. AJW thanks  the  Science  and  Technology  Facilities  Council  (STFC)  for  their  studentship. This work has been supported by the DISCSIM project, grant agreement 341137 funded by the European Research Council under ERC-2013-ADG. It has also used the DIRAC Shared Memory Processing system at the University of Cambridge, operated by the COSMOS Project at the Department of Applied Mathematics and Theoretical Physics on behalf of the STFC DiRAC HPC Facility (www.dirac.ac.uk). This equipment was funded by BIS National E-infrastructure capital grant ST/J005673/1, STFC capital grant ST/H008586/1, and STFC DiRAC Operations grant ST/K00333X/1. DiRAC is part of the National E-Infrastructure. This work has made use of data from the European Space Agency (ESA) mission {\it Gaia} (\url{https://www.cosmos.esa.int/gaia}), processed by the {\it Gaia} Data Processing and Analysis Consortium (DPAC, \url{https://www.cosmos.esa.int/web/gaia/dpac/consortium}). Funding
for the DPAC has been provided by national institutions, in particular the institutions participating in the {\it Gaia} Multilateral Agreement.

\bibliography{References/HVDO}{}
\bibliographystyle{mn2e}
\end{document}